\documentclass[aps,prb,final,twocolumn,superscriptaddress,floatfix,showpacs,10pt]{revtex4-1}

\usepackage[latin9]{inputenc}
\usepackage{graphicx,amssymb,soul,color,amsmath,bm}
\usepackage{epstopdf,hyperref}

\hypersetup{colorlinks=true,citecolor=blue}

\begin{document}

\title{
Photoexcitation of electronic instabilities in one-dimensional charge-transfer systems
}

\author{Juli\'an Rinc\'on}
\affiliation{Center for Nanophase Materials Sciences, Oak Ridge National Laboratory, Oak Ridge, Tennessee 37831, USA}

\author{K. A. Al-Hassanieh}
\affiliation{Center for Nanophase Materials Sciences, Oak Ridge National Laboratory, Oak Ridge, Tennessee 37831, USA}

\author{Adrian E. Feiguin}
\affiliation{Department of Physics, Northeastern University, Boston, Massachusetts 02115, USA}

\author{Elbio Dagotto}
\affiliation{Department of Physics and Astronomy, The University of Tennessee, Knoxville, Tennessee 37996, USA}
\affiliation{Materials Science and Technology Division, Oak Ridge National Laboratory, Oak Ridge, Tennessee 37831, USA}

\date{\today}

\begin{abstract}
We investigate the real-time dynamics of photoexcited electronic instabilities in a charge-transfer system model, using the time-dependent density matrix renormalization group method. The model of choice was the quarter-filled one-dimensional extended Peierls-Hubbard Hamiltonian interacting with classical few-cycle electromagnetic radiation. The results show that only one electronic instability drives the main features of the photogenerated time-dependent behavior. Indeed, the photoresponse of the system shows a large enhancement of the $4k_F$ (bond and charge) instability whereas the $2k_F$ state remains largely unaffected. This conclusion holds regardless of the nature of the optical excitations and whether the system is perturbed resonantly or not. Our results suggest potential applications of charge-transfer systems with slow phononic dynamics as optoelectronic switching devices.
\end{abstract}


\maketitle

\section{Introduction}
Accomplishing control over ultrafast and intense light pulses is at the 
heart of attosecond and femtosecond spectroscopy~\cite{Corkum07,Goulielmakis07}. 
Applications include subfemtosecond emission of extreme ultraviolet radiation, 
molecular dissociation, and the manipulation of changes in 
the structure of molecular, atomic, and solid state matter, just to mention 
a few~\cite{Corkum07,Orenstein12,Kampfrath13}. Manipulating such properties of materials via
light has opened a realistic and reliable route to the possibility of studying selected 
emergent states in complex systems such as superconductors, organic charge-transfer solids, 
Mott and charge density wave insulators, and others~\cite{Orenstein12,Kampfrath13}. 
Using resonant and intense nonresonant ultrafast pulses of 
electromagnetic radiation, remarkable 
experimental outcomes have been observed. Excellent examples include: photoinduced 
phase transitions, real-space scanning of molecular orbitals, control over dissociation 
of molecules, study of ionic and electronic motion, melting of ordered states like in superconductors, 
as well as the analysis of magnetic and charge order~\cite{Corkum07,Orenstein12,Kampfrath13}.

In the context of nonequilibrium properties, ultrafast optical pulses 
have allowed the excitation of electrons well above any intrinsic characteristic energy 
scale or manipulation of the degree of competition between different orders, since usually 
there is a complex interplay and exchange of energy between different many-body states 
of the system's Hilbert space. By probing the system's dynamics by 
shaking it with a pulse of light, it is possible to access states present 
in the energy spectrum which are not accessible via experiments carried out by 
changing the temperature~\cite{Chollet05,Onda08,Orenstein12}. These types 
of \emph{nonthermal states} often contain coexisting orders that are not 
usually present in the standard ground or thermal states, making a remarkable 
difference in the way the system evolves after photoexcitation. Experimental 
access to these highly nontrivial states has been accomplished in a 
variety of systems, as discussed above.

Quantum materials, also known as strongly correlated 
systems~\cite{Orenstein12}, belong to a family of systems where 
several phases compete producing extremely complex many-body states. 
Although a single phase may dominate the ground state, competing 
instabilities are often hidden at higher energies, which can be 
accessible with intense ultrafast pulses of light. The electromagnetic 
radiation will reshape the energy distribution of the many-body spectrum, 
reordering the relevance of the states, and in the case of a time-dependent 
electric field, switching rapidly from one to another~\cite{Orenstein12,Kampfrath13}.

When quantum materials are pumped with ultrafast light pulses, interesting dynamics 
ensues between several electronic instabilities such as charge order (CO), 
charge density waves (CDWs), spin density waves (SDWs), spin-Peierls (SP) 
states, bond order waves (BOWs), and others~\cite{Orenstein12}.~Representative 
materials that display several of these instabilities are the so-called 
charge-transfer salts: (TM)X$_2$, where TM is either TMTSF 
(tetramethytetraselenafulvalene) or TMTTF (tetramethyltetrathiafulvalene) 
while X$_2$ can be ClO$_4$, PF$_6$, or Br~\cite{Benthien05}. These compounds 
have in common that the filling factor can be considered as quarter-filled, 
in terms of either electrons or holes, and effectively they behave as one-dimensional 
(1D) chains. Typical instabilities found in these 
molecular systems are $2k_F$ and $4k_F$ BOW, CDW, CO, and SP states 
($k_F$ is the Fermi wavevector). Also important is the relative phase 
between the $2k_F$ and $4k_F$ instabilities which will define different 
ordered states (see Sec.~\ref{sec:proc}).

Using pump and probe ultrafast spectroscopy in the organic 
salt (EDO-TTF)$_2$PF$_6$, studies of the photoresponse of 
the optical conductivity, $\sigma(\omega)$, have reported a phase 
transition between two different CO states with a gigantic 
response in $\sigma(\omega)$~\cite{Chollet05,Onda08}. It was 
also shown that the photoinduced ordered state could not be 
assigned to any of the states of the thermal equilibrium 
spectrum, but rather to a nonthermal state~\cite{Chollet05,Onda08}. 
Similar experimental pump-and-probe studies of the photoexcited 
charge dynamics have been carried out on similar organic compounds~\cite{Okamoto07}.

Theoretical investigations of the photogenerated 
dynamics of organic compounds have also been 
reported~\cite{Maeshima05,Maeshima06,Maeshima07,Yonemitsu07,Yonemitsu09,Lee09,Moriya12,Uemura12,Lu12,Matsueda12}. Using a time-dependent Lanczos approach, the dynamics 
of (EDO-TTF)$_2$PF$_6$ induced by a time-dependent multi cycle 
electrical field studied the efficiency of the partial melting 
of the $2k_F$ CO ground state in the presence of different lattice potentials~\cite{Yonemitsu07}. 
That work used a quarter-filled 1D Hubbard model with Peierls and Holstein type 
of electron-phonon couplings, where the vibrational degrees of freedom were 
treated classically~\cite{Yonemitsu07}. Another related work, on a similar 
model and using the same numerical method, observed a complete melting of 
the CO, where the excess of energy associated with the order was transferred 
to the generation of phonons~\cite{Lee09}. After this process, a complete 
nonadiabatic decoupling between vibrational and electronic degrees of freedom 
was reported~\cite{Lee09}. Similar results for the half-filled 
case were reported elsewhere~\cite{Matsueda12}.

In this publication, we present a detailed density-matrix renormalization-group 
(DMRG)~\cite{dmrg1,dmrg2,dmrg3} study of the effect of few-cycle light pulses 
on the competition between charge and bond instabilities, present in organic 
compounds, in the pump-and-probe situation. We resort to a quarter-filled 
1D extended Peierls-Hubbard Hamiltonian to model the molecular compounds 
interacting with the incident radiation. This model mimics the setup used 
in ultrafast spectroscopy measurements, where a portion of the material is pumped 
with an ultrafast light pulse, lasting typically a few femtoseconds. %
Here, we focus on the photoresponse of the charge and bond instabilities, 
and how the system reacts to the interaction with both resonant and nonresonant 
electric field. Moreover, we also explore all the possible optical excitations 
of the system, namely, fermionic optical excitations, holon-antiholon pairs, 
as well as excitons.
Our main finding is that the states that dominate the real-time 
dynamics largely correspond to oscillating $4k_F$ charge and bond instabilities; 
however, some intermediate states that were observed during the evolution 
of the system are dominated by the $2k_F$ instability. These results hold 
for both resonant and nonresonant radiation. We also discuss the relevance 
of our results to recent experiential findings 
in organic quasi-1D organic salts.%
Our numerical results, indicating an ultrafast switching from $2k_F$ to $4k_F$ states, 
compare well with prior reports that showed a partial or complete melting of the $2k_F$ 
ground state~\cite{Yonemitsu07,Lee09}.

The outline of the paper is the following: in Sec.~\ref{sec:model} 
we discuss the Hamiltonian model and the optical excitations relevant 
to the photoinduced dynamics; Sec.~\ref{sec:proc} focuses on the 
numerical method and procedures used to implement the electric field; 
in Sec.~\ref{sec:res} we present the results for the time-dependent 
behavior of the electronic instabilities and the corresponding 
analysis; and finally, in Sec.~\ref{sec:con} we close with the conclusions.

\section{Model and optical excitations\label{sec:model}}
In this section, we describe the model Hamiltonian used in the study 
of the light-induced electronic instabilities resulting from the 
electron-electron and electron-phonon interactions. The electronic 
degrees of freedom are accounted for using the 1D extended Hubbard 
Hamiltonian which includes local and nearest-neighbor repulsion between 
particles. The electron-phonon interaction is incorporated via the Peierls 
coupling: a dimerization term with a frozen lattice distortion. 
The phonons are considered to have a much slower dynamics than 
that of the electrons, i.e., phonons with no dynamics. The resulting 
extended Peierls-Hubbard (EPH) 
Hamiltonian~\cite{Ung94,Riera00,Shibata01,Tsuchiizu01,Kuwabara03,Benthien05} reads
%
\begin{equation}
\begin{aligned}
H = &-\sum_{i,\sigma}\left(t_{i,i+1}\, c^+_{i,\sigma} c_{i+1,\sigma} + \mathrm{H.c.}\right) + U\sum_{i} n_{i,\uparrow} n_{i,\downarrow} \\&+ V\sum_{i} \left(n_{i}-n\right)\left(n_{i+1}-n\right),
\end{aligned}
\label{eq:H}
\end{equation}
where $t_{i,i+1}=t(1-(-1)^i\delta/2)$ represents the electron-phonon 
coupling via the dimerization term $\delta$, and $t$ is the electronic 
hopping between neighboring sites in the absence of the dimerization. 
The parameters $U$ and $V$ are the local and nearest-neighbor Coulomb 
repulsion, respectively; the rest of the notation is 
standard~\cite{Ung94,Riera00,Shibata01,Tsuchiizu01,Kuwabara03,Benthien05}. 
The filling is set to quarter filling, $n=1/2$, and can be 
interpreted as electrons or holes depending on the specific organic compound under consideration.

Since we are only interested in the electronic properties, we neglect 
any dynamical terms related to the light and consider it as an external 
classical field. The interaction with light is incorporated via the minimal 
coupling, which when using the flux or velocity gauge~\cite{Madsen02}, 
is expressed in second quantization by an effective flux that modifies 
the hopping term as $t\rightarrow t\,e^{i \phi/L}$, where $L$ is the 
system's length. The magnetic flux, $\phi$, comes from the oscillating 
vector potential dependent on time, $\tau$: $\phi = a\,A(\tau)$ with $a$ 
the lattice constant and where
\begin{equation}
A(\tau) = A_0\, e^{-(\tau-\tau_p)^2/2\sigma^2}\cos\left[\omega_p(\tau-\tau_p)\right],
\label{eq:A}
\end{equation}
is the explicit form of the vector potential~\cite{Madsen02}. $A_0$, $\omega_p$, and $\sigma$ represent the intensity, frequency, and width of the light pulse and $\tau_p$ corresponds to the time when the electric field reaches its maximum value.

We now revisit some of the properties of the EPH model relevant to our study. The EPH model displays a variety of ground states depending on the values of $U$, $V$, and $\delta$. Usually, ground and excited states include coexistence of $2k_F$ and $4k_F$ CO, CDW, BOW, SP, and SDW states~\cite{Ung94,Riera00,Shibata01,Kuwabara03}. At quarter filling, the ground state is an insulator that is driven by the opening of the dimerization gap $\Delta=2\delta t$ and the effect of the interactions $U$ and $V$~\cite{Tsuchiizu01}. 
In principle, all kinds of combinations of instabilities are possible; 
however, those that are truly relevant will depend on the type of experiment 
and material under scrutiny. Different materials properties are driven by 
different types of interactions (this is reflected in the values taken by 
the parameters of the EPH Hamiltonian). Therefore, different competing 
instabilities will form the excitation spectrum and, accordingly, different 
states will be sampled in a pump-and-probe experimental setup.


Also important for our purposes are the optical excitations of 
the EPH model~\cite{Benthien05}. In the limit of a large dimerization gap, $\delta \lesssim 2$ and $V=0$, optical excitations will be formed by a pair of fermionic quasi-particles with opposite spin and charge: one hole in the valence band and one electron in the conduction band. Whereas for $V<2t$ and $\delta <1$, the optical excitations are made of unbound spinless holon-antiholon pairs: the holon (antiholon) belongs to the lower (upper) Hubbard. For $V>2t$ and $\delta <1$, these unbound pairs bond, forming Mott-Hubbard excitons.

In the limit of large dimerization, the optical conductivity has three clear spectral structures. (1) For energies, $\omega$, larger than the dimerization gap, the optical spectrum has contributions from interdimer excitations with peaks at $\omega=2t(1+\delta/2)$ and $\omega=2t(1+\delta/2)+U$. These energies correspond to the annihilation of a bonding state in one dimer and the creation of an antibonding state on the neighboring dimer. In particular, $\omega=2t(1+\delta/2)$ corresponds to the formation of a triplet state on the second dimer giving rise to spin-Peierls coupling. (2) Below the dimerization gap, the optical excitations are those of an effective half-filled Hubbard chain with an absorption band centered around $\omega=U_{\rm eff}=U/2$. (3) For $\omega=\Delta$, there is a narrow absorption band associated with intradimer excitations of bandwidth proportional to $t(1-\delta/2)$. Most of the optical weight is located at this energy~\cite{Benthien05}.

Increasing interactions and decreasing the dimerization lead to a 
different structure of the optical spectrum. The optical conductivity 
is made of one absorption band (a continuum of unbound holon-antiholon pairs) 
that starts at the Mott gap and has a maximum close to it. The position of the 
maximum and the onset of the spectral weight depend on the interaction parameters. 
This maximum will become the Drude peak in the limit $\delta\rightarrow 0$; spectral 
features around $\omega=U$ are also expected and are related to optical excitations 
from the lower to the upper Hubbard bands~\cite{Jeckelmann04,Benthien05}. 

As in the case of $U=0$, the optical gap equals the Mott gap for $V<2t$. 
Therefore, the low-energy spectrum still corresponds to unbound pairs of 
charged excitations. For larger $V>2t$, the optical gap is smaller than 
the Mott gap signaling the appearance of an excitonic peak, radically 
changing the low-energy spectrum~\cite{Jeckelmann04,Benthien05}.

The presence of the time-dependent electric field will generate 
inter- and intradimer excitations, excitons, or holon-antiholon 
pairs at different characteristic timescales, which are associated 
with the energy of these excitations, depending on the parameters 
of the EPH model. Therefore, the understanding of the optical excitations 
in the system will prove crucial to the understanding of the photoinduced 
dynamics and concomitant melting of the charge and bond orders. 
We will see below that the photoinduced dynamics is dominated by the 
aforementioned optical excitations.

\section{Method and procedures\label{sec:proc}}
We now describe the procedure followed in the calculation of the photoinduced dynamics. The ground sate of the EPH model, i.e.~for $A=0$, was calculated using static DMRG~\cite{dmrg1,dmrg2,dmrg3}. Then, using time-dependent DMRG (t-DMRG)~\cite{White04,Daley04,dmrg2,dmrg3}, we applied the light pulse for a time interval $t_{\rm pump}=8\sigma$ with $A(\tau)$ as given in Eq.~(\ref{eq:A}). 
Once the pulse was applied, the system is subsequently time evolved 
under $H$ with $A=0$, i.e., the static EPH model [Eq.~(\ref{eq:H})], 
using again t-DMRG. During the entire time evolution, we calculated 
the time dependence of the mean value of charge, $\langle n_i\rangle$, 
and the correlations associated with bond $\langle c^+_{i+1}c_i\rangle$, 
charge $\langle n_{i+1}n_i\rangle$, and spin $\langle S^+_{i+1}S^-_i\rangle$ orders. 

In order to calculate the weight of the electronic instabilities 
from the t-DMRG results, we have performed fittings~\cite{Note1} 
for each time slice to the following parametrizations of the 
electronic instabilities~\cite{Riera00,Kuwabara03}
\begin{equation}
\begin{aligned}
\Delta n_i&=n_{4k_F}\cos(4k_Fr_i)+n_{2k_F}\cos(2k_Fr_i+\Phi_{2k_F}), \\
b_i&=b_{4k_F}\cos(4k_Fr_i)+b_{2k_F}\cos(2k_Fr_i+\Phi_{2k_F}),
\label{eq:ins}
\end{aligned}
\end{equation}
for charge ($\Delta n_i:=n_i-n$), spin, and bond orders; here, $r_i/a=i$, the Fermi wavevector is $k_F=n\pi/2$, and $c_{4k_F}$ and $c_{2k_F}$ characterize the amplitude of the modulations of the $4k_F$ and $2k_F$ excitations ($c=n$ and $b$ for charge and bond orders respectively). A relative phase, $\Phi_{2k_F}$, has been included to account for both states, site- and bond-centered waves~\cite{Riera00}. Notice that this phase takes commensurate values when describing different ordered ground states; however, we will see that $\Phi_{2k_F}$ can also take incommensurate values for the states that arise from 
the photostimulation. In order to understand the time-dependent 
tendency of the photoexcited instabilities, a Fourier transform 
analysis was performed and compared to the optical excitations.

The reliability of the value of the instabilities coming from the fittings have been cross-checked with calculations of order parameters that are obtained 
from the time-dependent correlation functions of the bond, charge, and spin orders, namely,
\begin{equation}
\begin{aligned}
\langle n_{4k_F}^{O}\rangle &= \frac{1}{L}\sum_i (-1)^i\langle n_i\rangle,\\
\langle n_{2k_F}^{O}\rangle &= \frac{1}{L}\sum_{i} (-1)^{\lfloor i/2\rfloor}\langle n_i\rangle,\\
\langle b^O\rangle &=\frac{1}{L}\sum_{i,\sigma} \langle c_{i+1\sigma}^+c_{i\sigma}+\mathrm{H.c.}\rangle.
\end{aligned}
\end{equation}

Our static and time-dependent DMRG simulations have been 
done for several system sizes, $L=12-48$ sites, while
up to $m=400$ states per block were kept leading to 
discarded weights of $10^{-6}$ for the longest time 
reached. The results shown here correspond to $L=24$ up to $\tau\approx 90/t$; however, similar $\tau$ dependence was observed for $L=36$ in the same time range, and for $L=48$ at early times. We notice that the finite-size effects in the EPH model are small for $L\gtrsim 20$, as previously report in Ref.~\onlinecite{Kuwabara03}. The time step was set to $\tau=0.05 ~[1/t]$, 
the runs were done up to times $\tau\sim 120 ~[1/t]$, and we have used a third-order Suzuki-Trotter expansion of the evolution operator. Open boundary conditions were imposed.

\section{Results\label{sec:res}}
In this section, we will discuss our main results corresponding to the 
time-dependent photoinduced dynamics of the electronic 
instabilities in the EPH model, Eq.~(\ref{eq:H}), as obtained 
with t-DMRG. The particular set of parameters chosen were 
used in previous dynamical DMRG calculations of the optical 
conductivity in the EPH model, allowing us to know in advance the main excitation 
frequencies~\cite{Benthien05}. First, we will discuss the case of 
interaction with resonant light, and then we will discuss the 
regime of a nonresonant intense electromagnetic perturbation.

\subsection{Resonant case}
\begin{figure}
\includegraphics*[width=.8\columnwidth]{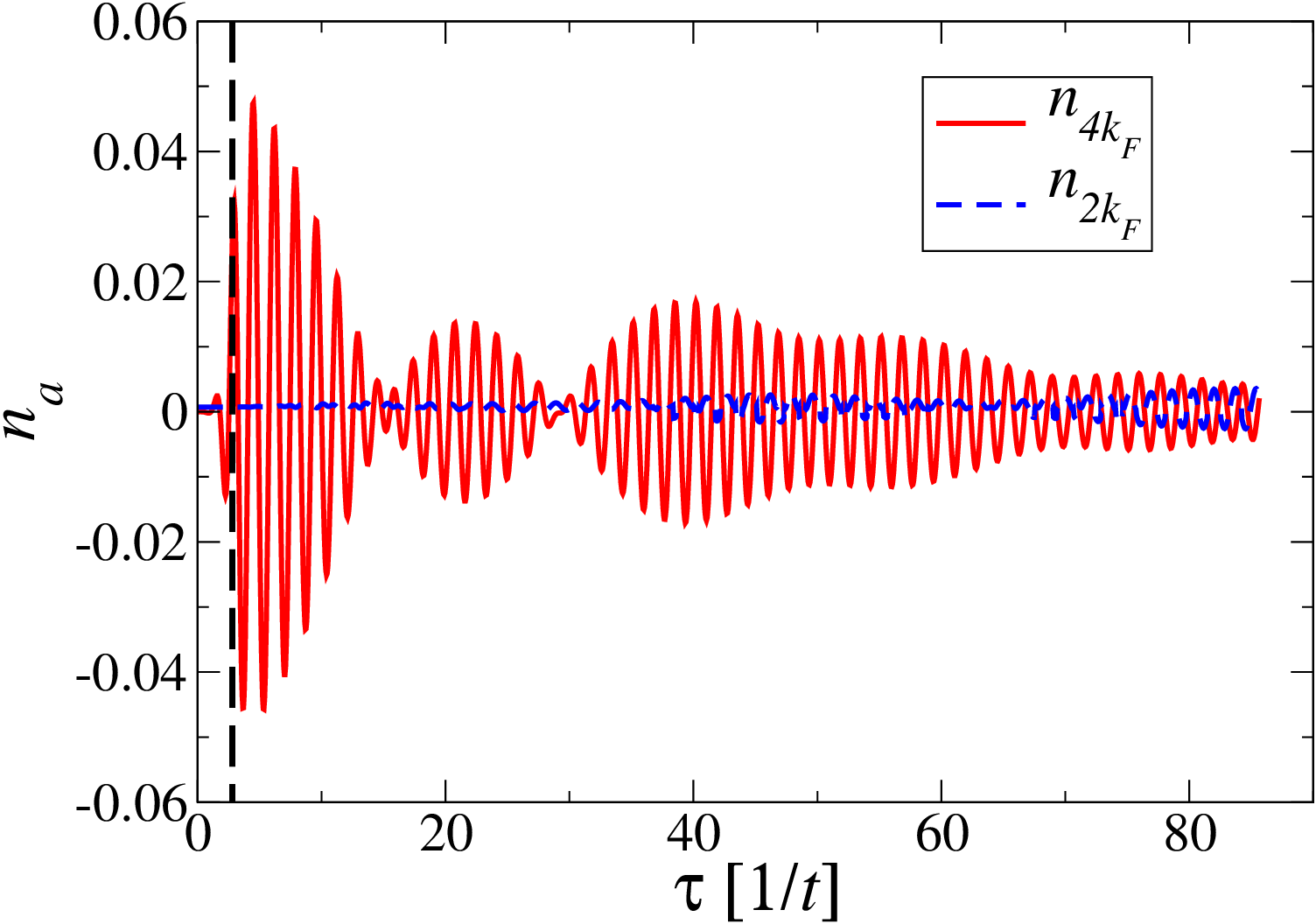}
\includegraphics*[width=.8\columnwidth]{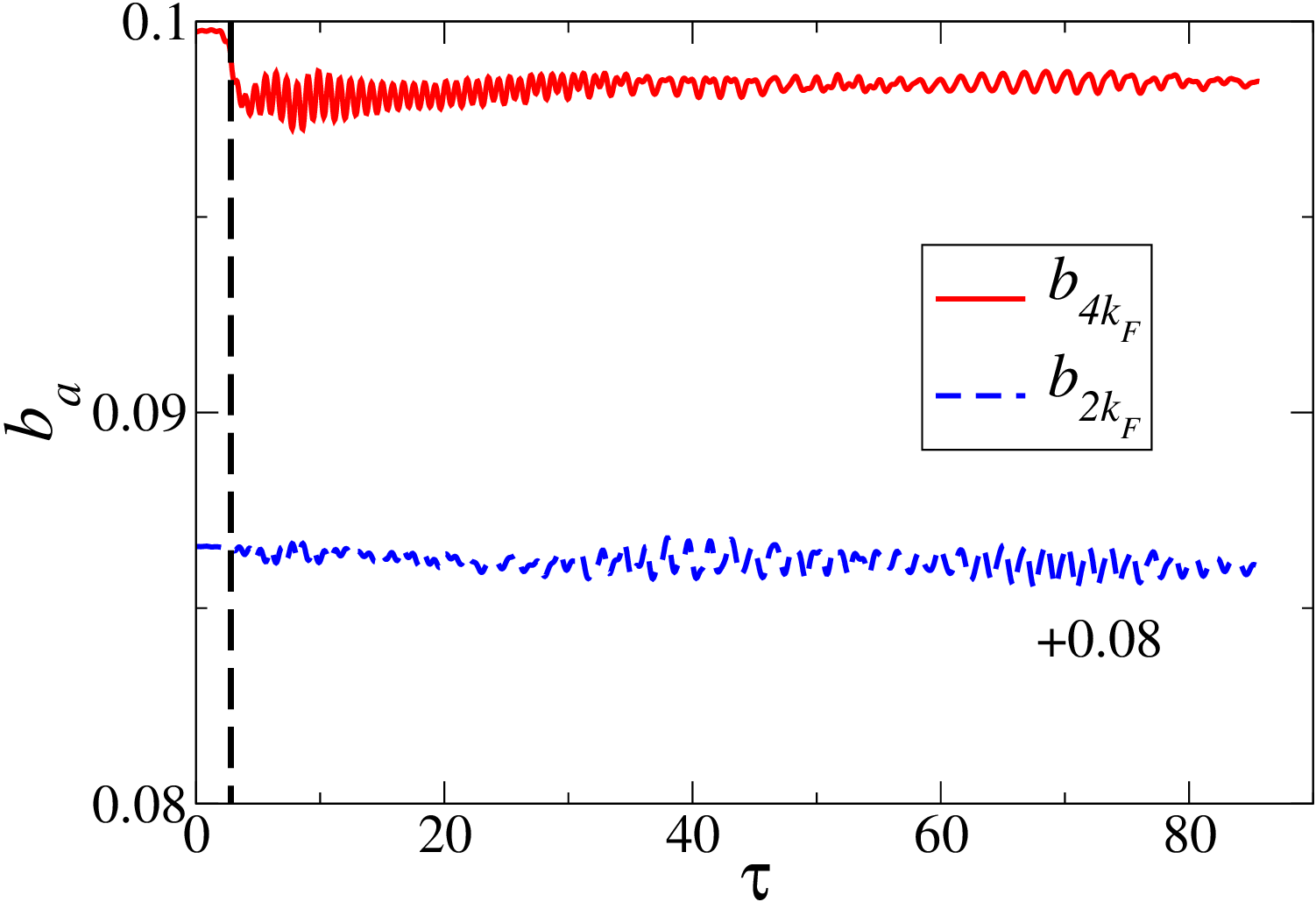}
\caption{Photoexcited time evolution of the charge (top) and bond (bottom) 
instabilities, $a=2k_F$ and $4k_F$, in the large dimerization limit. 
The vertical dashed line corresponds to the maximum amplitude of the 
resonant light at $\tau_p$ and parameters: $\delta=1.64$, $U=3.64t$, 
$V=0$, $A_0=1.75$, $\omega_{p}=4.36t$, $\sigma=0.7/t$.}
\label{fig:1}
\end{figure}

Let us start by analyzing the results shown in Fig.~\ref{fig:1}, 
considering the discussion of Sec.~\ref{sec:proc}, in the limit of 
large dimerization. The parameters chosen were $\delta=1.64$, 
$U=3.64t$, $V=0$~\cite{Benthien05}; the frequency and amplitude 
of the electric field were $\omega_p=4.36t$, $\sigma=0.7/t$, and 
$A_0=1.75$. In this case the singularities in $\sigma(\omega)$ are 
square-root divergences and, therefore, clear oscillations are expected 
in the photodynamics~\cite{Jeckelmann04,Benthien05}. The top plot in 
Fig.~\ref{fig:1}, shows the instabilities associated with the charge. 
At $\tau=0$, the dominant state is the $2k_F$ instability: $n_{2k_F}\approx 0.001$ 
and $n_{4k_F}\approx 10^{-7}$. Although $n_{2k_F}$ is small, there is a clear tendency towards this instability when inspecting $n_i$ vs $i$ (not shown). We notice that $n_{2k_F}$ is finite due to the open boundaries; nevertheless, considering such boundaries as scattering impurities, it is interesting to explore its role in the relaxation and effect of the photodynamics of the system. As the system is being pumped, the $4k_F$ instability 
is greatly enhanced and shows clear oscillations that can be related to the 
intra-dimer band present in the $\sigma(\omega)$ at $\omega=2t(1+\delta/2)$. 
Inter-dimer excitations at the same energy also contribute to the dynamics; 
as discussed above, this excitation energy, at $\omega=2t(1+\delta/2)$, is 
also associated with spin degrees of freedom. The frequency of the oscillations 
corresponds to energies associated with the edges and center of this band. An 
overall decay of $n_{4k_F}$ and a slight increase of $n_{2k_F}$ 
is observed as time advances, eventually leading to an asymptotic 
state which resembles the original ground state.

The bottom plot in Fig.~\ref{fig:1} shows the time evolution of the 
bond instabilities in the large dimerization limit. For $\tau=0$, the 
ground state possesses an SP coupling, as corroborated by the quantities 
$\langle S^+_{i+1}S^-_i\rangle$ and $\langle b^O\rangle$ calculated with 
static DMRG, with contributions from both $2k_F$ and $4k_F$ instabilities. 
After the system is pumped, we see a decrease in $b_{4k_F}$ and a subsequent
 recovery. The small changes in the bond order can be related to the existence 
of strong dimers due to the large value of $\delta$. At early times the 
frequency of oscillations corresponds to the energy $\omega=2t(1+\delta/2)+U$; 
on the other hand, $b_{2k_F}$ alternates with a frequency $\omega=2t(1+\delta/2)$ 
which is the energy of the formation of the triplet state, signaling the 
presence of the SP state. A spectral analysis (not shown)~\cite{Note2} also shows the generation of 
holon-antiholon pairs at approximately the energy $\omega=U_{\rm eff}/2$ 
related to the effective Hubbard chain (see Sec.~\ref{sec:model}).

\begin{figure}
\includegraphics*[width=.8\columnwidth]{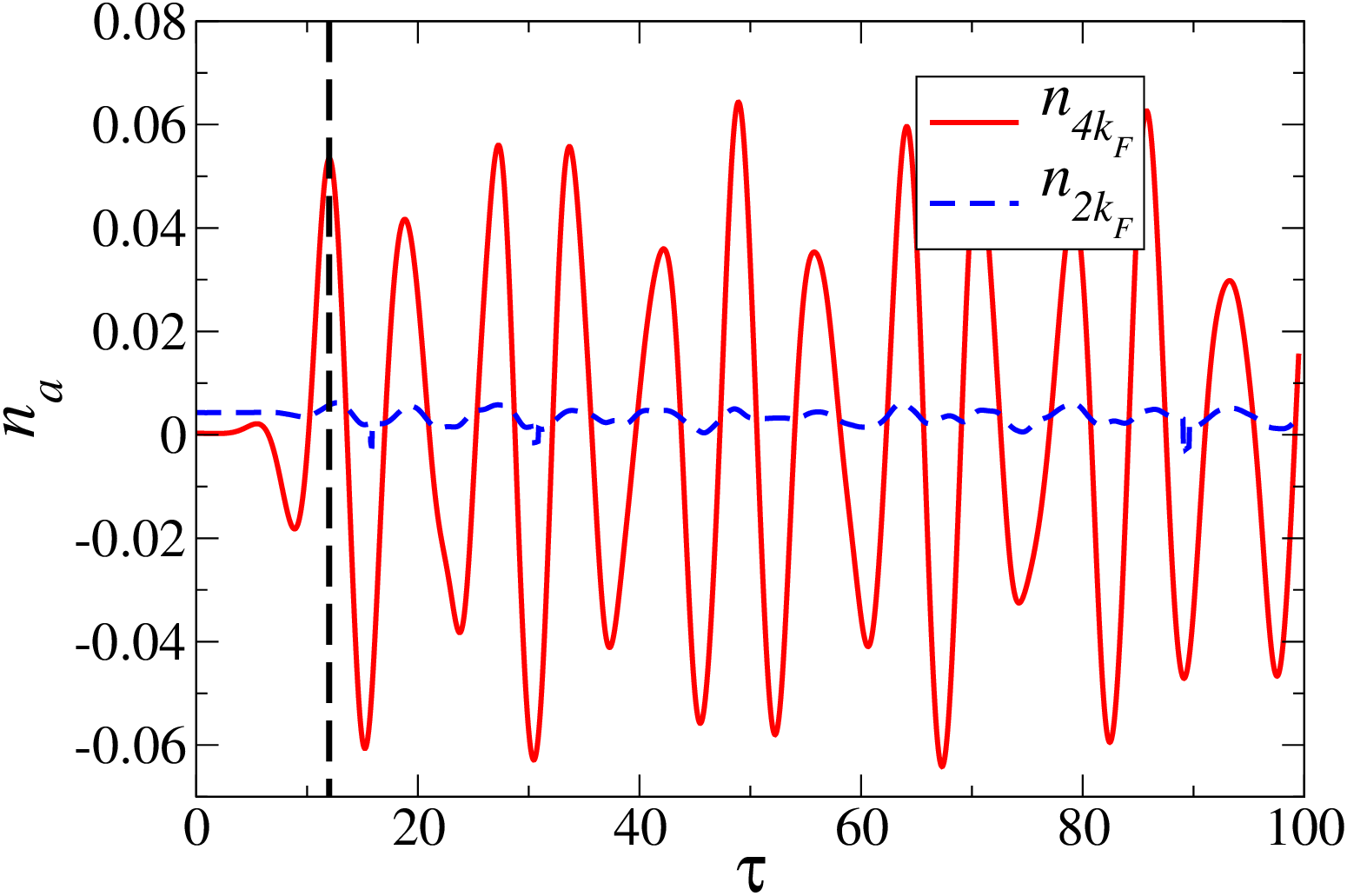}
\includegraphics*[width=.8\columnwidth]{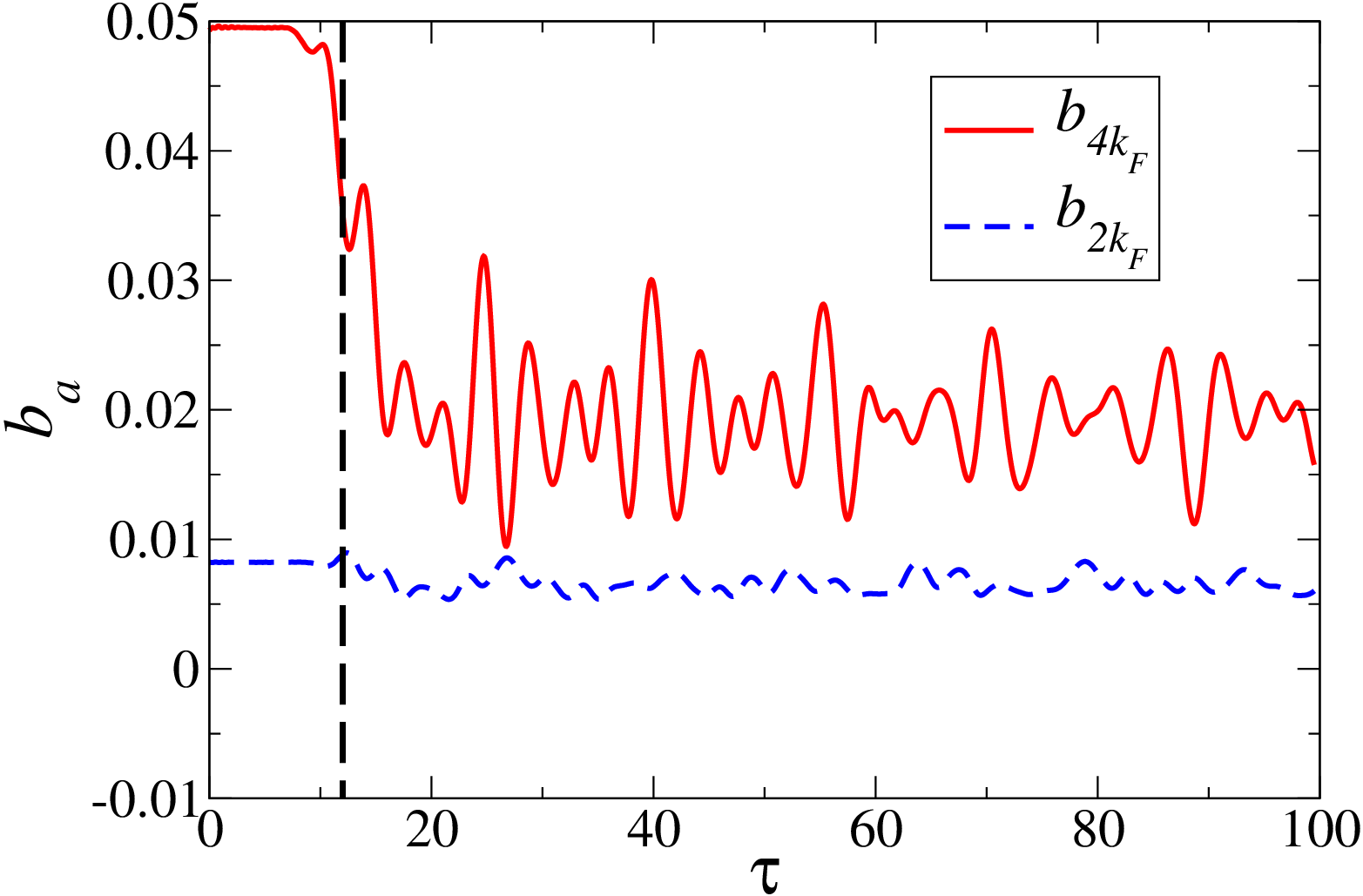}
\caption{Photoexcited time evolution of the charge (top) and bond (bottom) instabilities, $a=2k_F$ and $4k_F$. The vertical dashed line corresponds to the maximum amplitude of the resonant light at $\tau_p$ and parameters: $\delta=0.35$, $U=8.24t$, $V=1.64t$, $A_0=2.75$, $\omega_{p}=0.81t$, $\sigma=3/t$.}
\label{fig:2}
\end{figure}

Decreasing $\delta$ and increasing $U$ leads to a square-root onset 
of spectral weight at the optical gap as the main feature in 
$\sigma(\omega)$~\cite{Jeckelmann04,Benthien05}. Therefore, in this case 
we do not expect sharp oscillations due to the lack of resonances (singularities). 
That is why we have set $\omega_p$ to the maximum that appears after the onset of 
spectral weight. Figure~\ref{fig:2} shows the photodynamics for the case $\delta=0.35$, 
$U=8.24t$, $V=1.64t$~\cite{Benthien05}; the electric field parameters are $A_0=2.75$, 
$\omega_{p}=0.81t$, $\sigma=3/t$. For this set of parameters the ground state 
corresponds to a $2k_F$ CDW with $\dots 1100 \dots$ CO plus a $2k_F$ and $4k_F$ BOW state~\cite{Ung94,Kuwabara03}. The charge 
instabilities (top plot, Fig.~\ref{fig:2}) display oscillations that can 
be associated with energies close to the edge of the optical absorption 
spectrum set by the Mott gap $\omega\sim E_c=0.95t$. As the electric field 
pumps the system, $n_{2k_F}$ gets reduced whereas $n_{4k_F}$ is enhanced; 
and as time evolves, only $n_{4k_F}$ contributes to the overall dynamics. 
Notice that the apparent on-phase behavior of $n_{2k_F}$ and $n_{4k_F}$ is 
corrected by the presence of the phase $\Phi_{2k_F}$, which oscillates nontrivially (not shown).

As for the time-dependent behavior of the bond instabilities (Fig.~\ref{fig:2}, bottom), both $b_{4k_F}$ and $b_{2k_F}$ are reduced as the system is excited by the external radiation. The enhanced reduction in $b_{4k_F}$ can be attributed to a strong coupling between the current operator and states with a unit cell of two sites. The  characteristic frequencies have the same physical nature as in the case of the charge; i.e., they belong to unbound optical excitations. Similar results for the photodynamics are found for the case $V=0$ (not shown). This type of trend in the evolution of the electronic instabilities is expected to be valid for $V<2t$, where no bound excitations (excitons) are created.


\begin{figure}
\includegraphics*[width=.8\columnwidth]{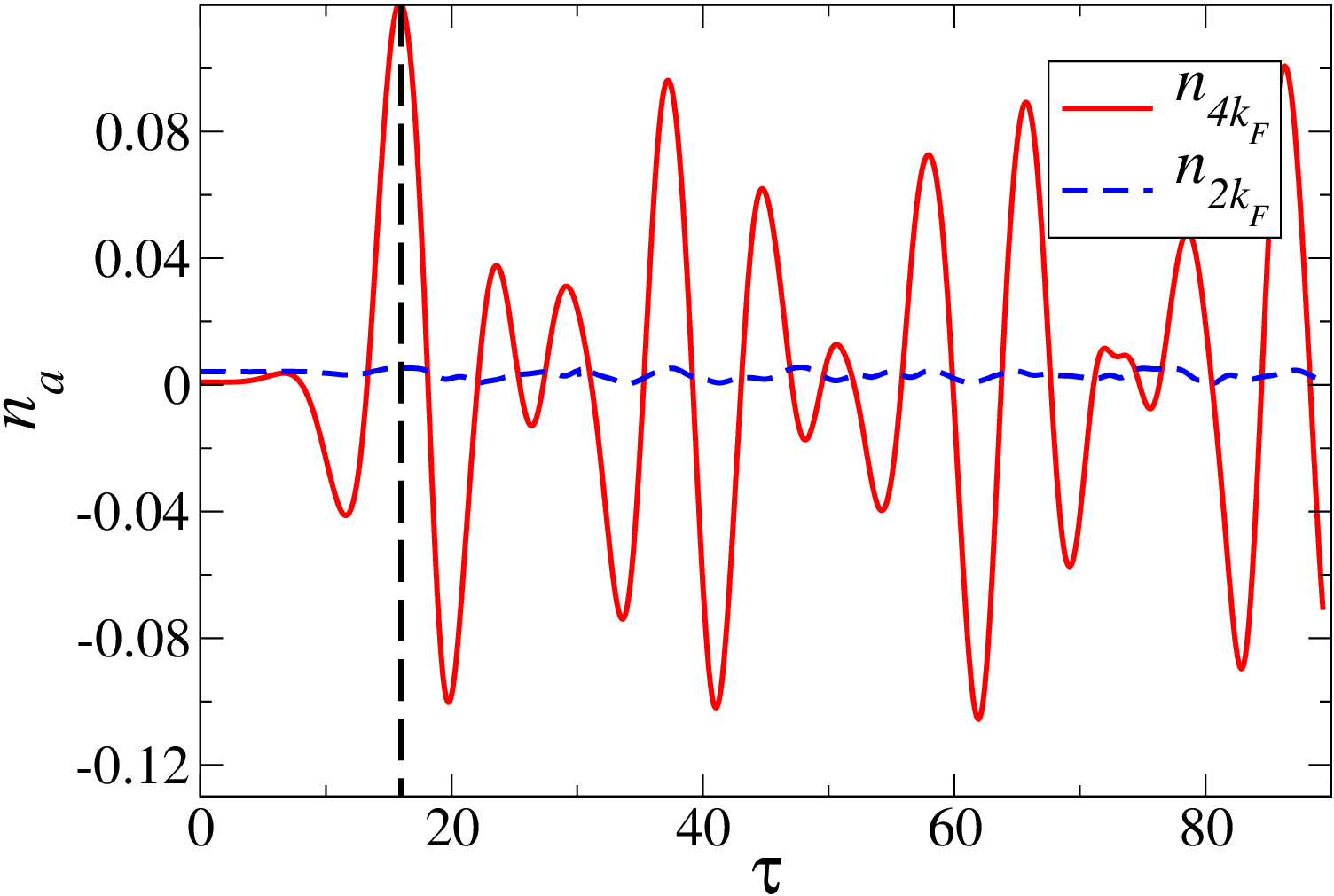}
\includegraphics*[width=.8\columnwidth]{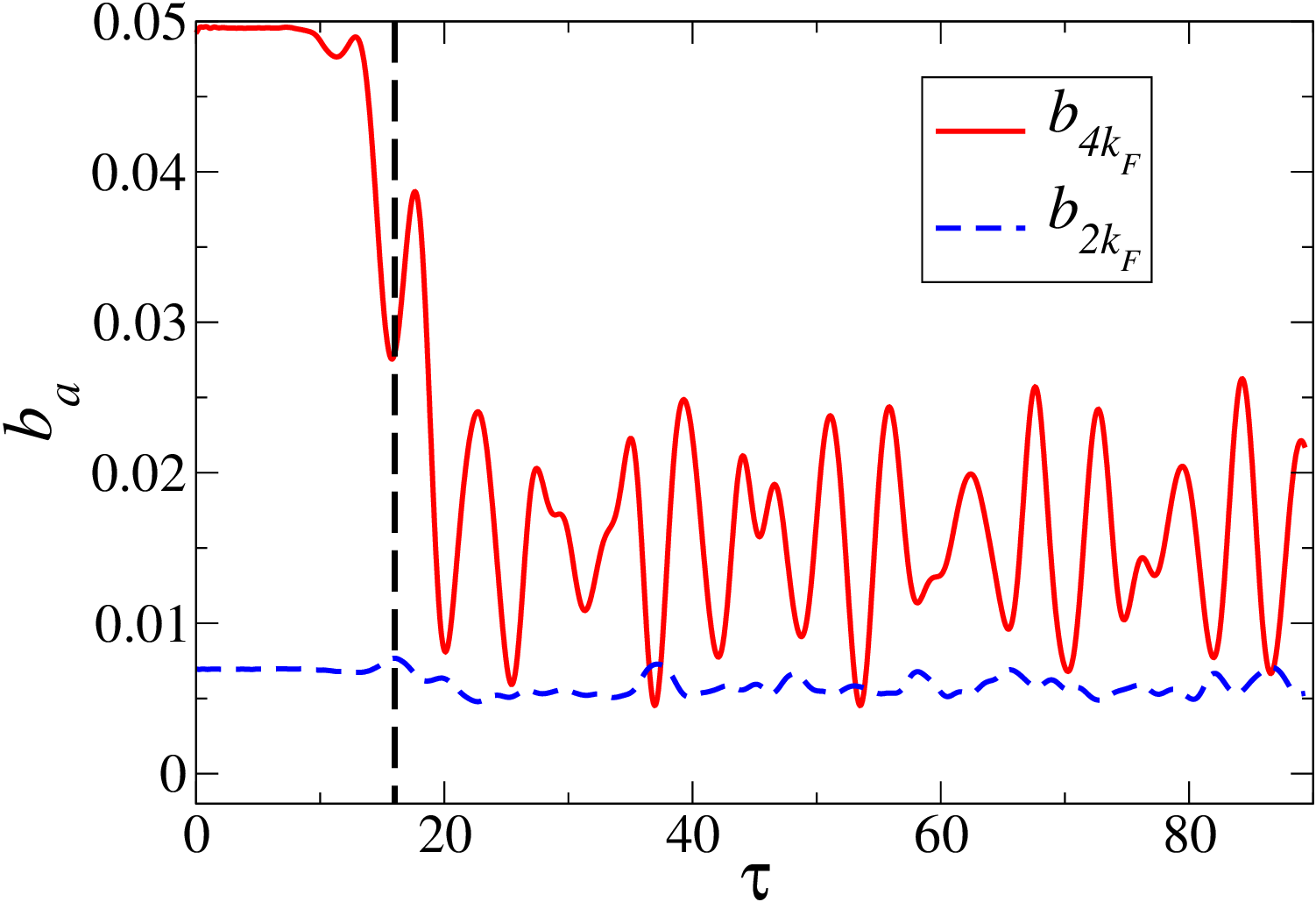}
\caption{Photoexcited time evolution of the charge (top) and bond (bottom) instabilities, $a=2k_F$ and $4k_F$. The vertical dashed line corresponds to the maximum amplitude of the resonant light at $\tau_p$ and parameters: $\delta=0.35$, $U=8.24t$, $V=3.29t$, $A_0=3.75$, $\omega_{p}=0.55t$, $\sigma=4/t$.}
\label{fig:3}
\end{figure}

As discussed in Sec.~\ref{sec:model}, for $V>2t$, the presence of excitons radically modify 
the spectral properties~\cite{Jeckelmann04,Benthien05}. A well-defined resonance (a delta peak) 
at the optical gap leads to sharp oscillations in the instabilities as a function of time. 
Figure~\ref{fig:3} shows the oscillations of the electronic instabilities for the case 
$\delta=0.35$, $U=8.24t$, $V=3.29t$~\cite{Benthien05}, and electric field parameters 
$A_0=3.75$, $\omega_{p}=0.55t$, $\sigma=4/t$, where the presence of excitons has been 
shown before~\cite{Benthien05}. The evolution in time of $n_{4k_F}$ (Fig.~\ref{fig:3}, top) 
shows an abrupt increase as the system is pumped, reaching its maximum around the radiation 
pulse maximum; whereas $n_{2k_F}$ is fairly insensitive to the radiation. Interestingly, 
the charge oscillation is related to the bound energy of the exciton $E_b=0.9t$ and the 
excitonic energy that equals the optical gap. The characteristic frequency comes from 
resonance between the excitonic peak at $\omega=0.55t$ and the onset of the absorption 
band of holon-antiholon pairs ($E_c=1.4t$).

The bond instabilities are shown in the bottom panel of Fig.~\ref{fig:3}. Similarly 
to the case of charge, in this case the photodynamics is dominated by $b_{4k_F}$ 
while $b_{2k_F}$ remains slightly unaffected, at least in the time domain studied. 
Similarly as in the situation in the previous figures, once the electric field reaches 
its maximum, $b_{4k_F}$ is considerably reduced and, at later times, there is a 
small recovery of the bond instability accompanied by oscillations related to 
the bound energy of the exciton.

We conclude this section noticing that the oscillations of the photogenerated 
dynamics are greatly modified by the optical excitations and by whether these excitations 
are true singularities or not. It is also important to notice that the phase 
$\Phi_{2k_F}/\pi$ takes oscillatory incommensurate values (not shown), giving 
rise to states that differ from the ground state situation where $\Phi_{2k_F}/\pi$ 
can only take commensurate values. Notice that these states 
can be associated with nonthermal states observed in spectroscopy experiments~\cite{Chollet05,Onda08}.

\begin{figure}
\includegraphics*[width=.8\columnwidth]{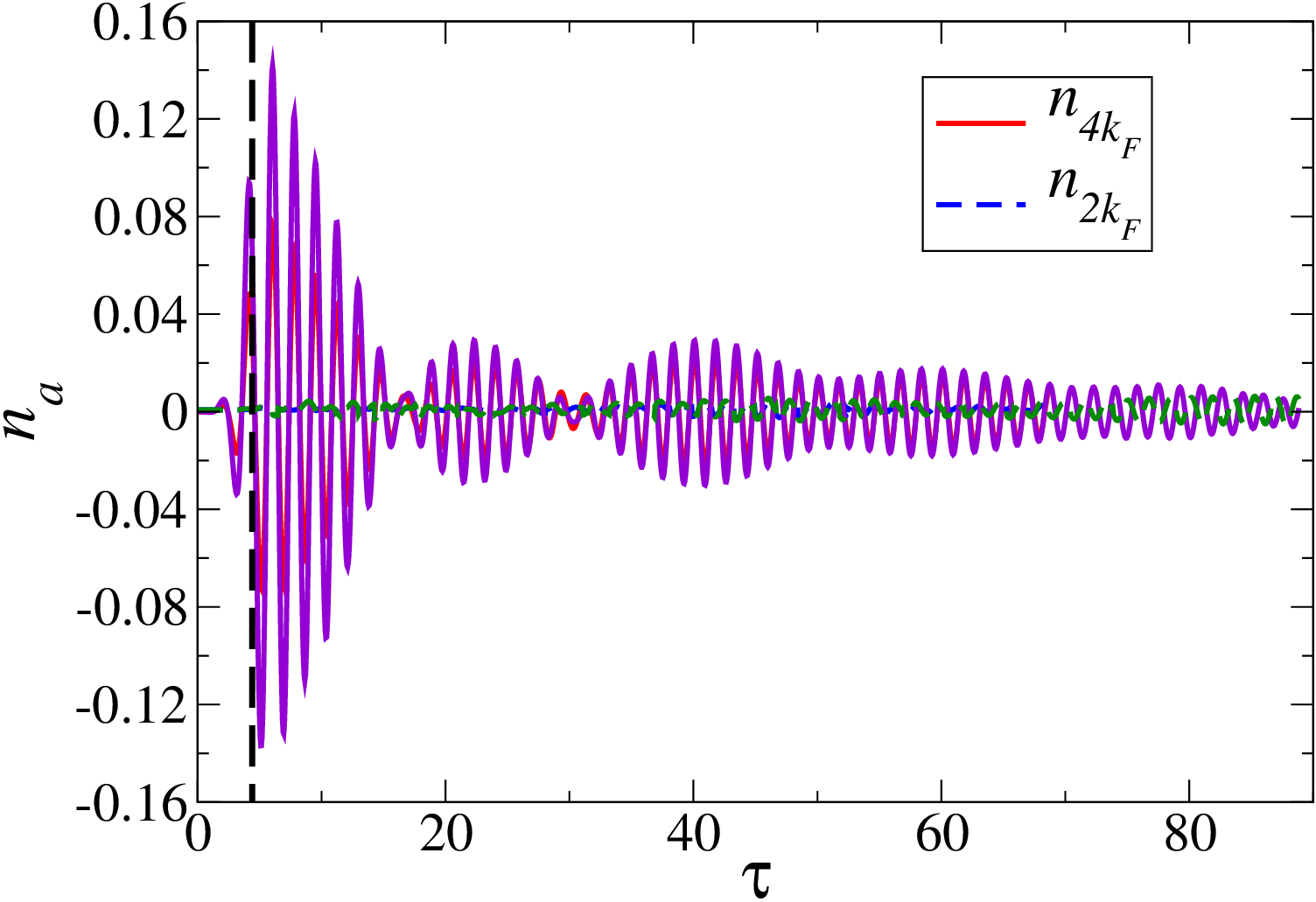}
\includegraphics*[width=.8\columnwidth]{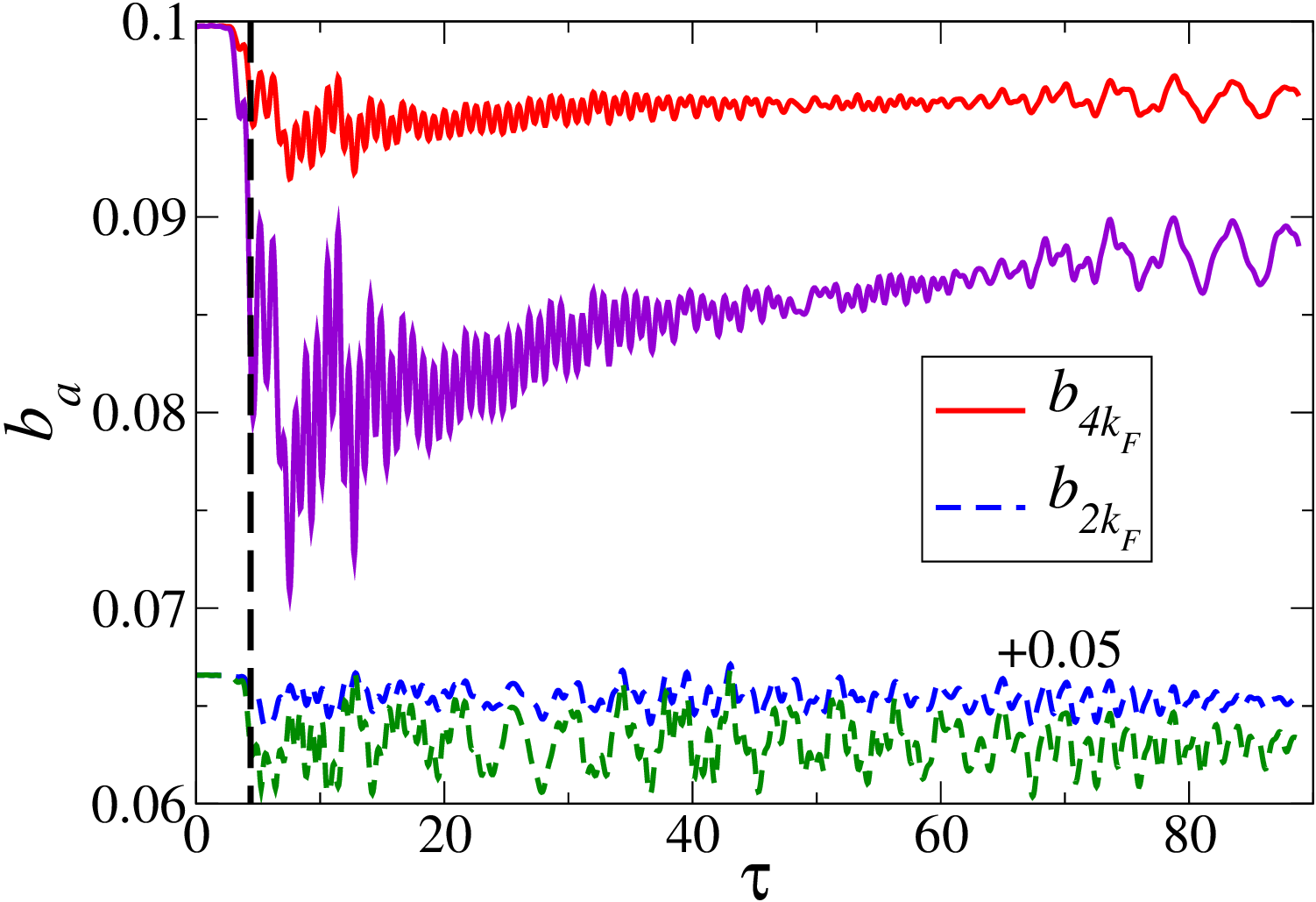}
\caption{Photoexcited time-evolution of the charge (top) and bond (bottom) instabilities, $a=2k_F$ and $4k_F$, in the large dimerization limit. The vertical dashed line corresponds to the maximum amplitude of the nonresonant light at $\tau_p$ and parameters: $\delta=1.64$, $U=3.64t$, $V=0$, $A_0=3.5$ (red and blue) and 7 (violet and green), $\omega_{p}=2.5t$, $\sigma=1.1/t$.}
\label{fig:4}
\end{figure}

\subsection{Nonresonant case}
In this section we will explore the effects of a nonresonant intense electric field which takes advantage of the ponderomotive effect. It is expected that this effect will give rise to similar results as in the resonant case. All the results shown in this section were done using a frequency of the electric field $\omega_p=2.5t$, $\tau_p=4.4/t$, and $\sigma=1.1/t$; we notice that this value does not match any optical excitations~\cite{Benthien05}. We also used the same parameters in the EPH model as in the resonant case so we can make a direct comparison with the nonresonant situation.

\begin{figure}
\includegraphics*[width=.8\columnwidth]{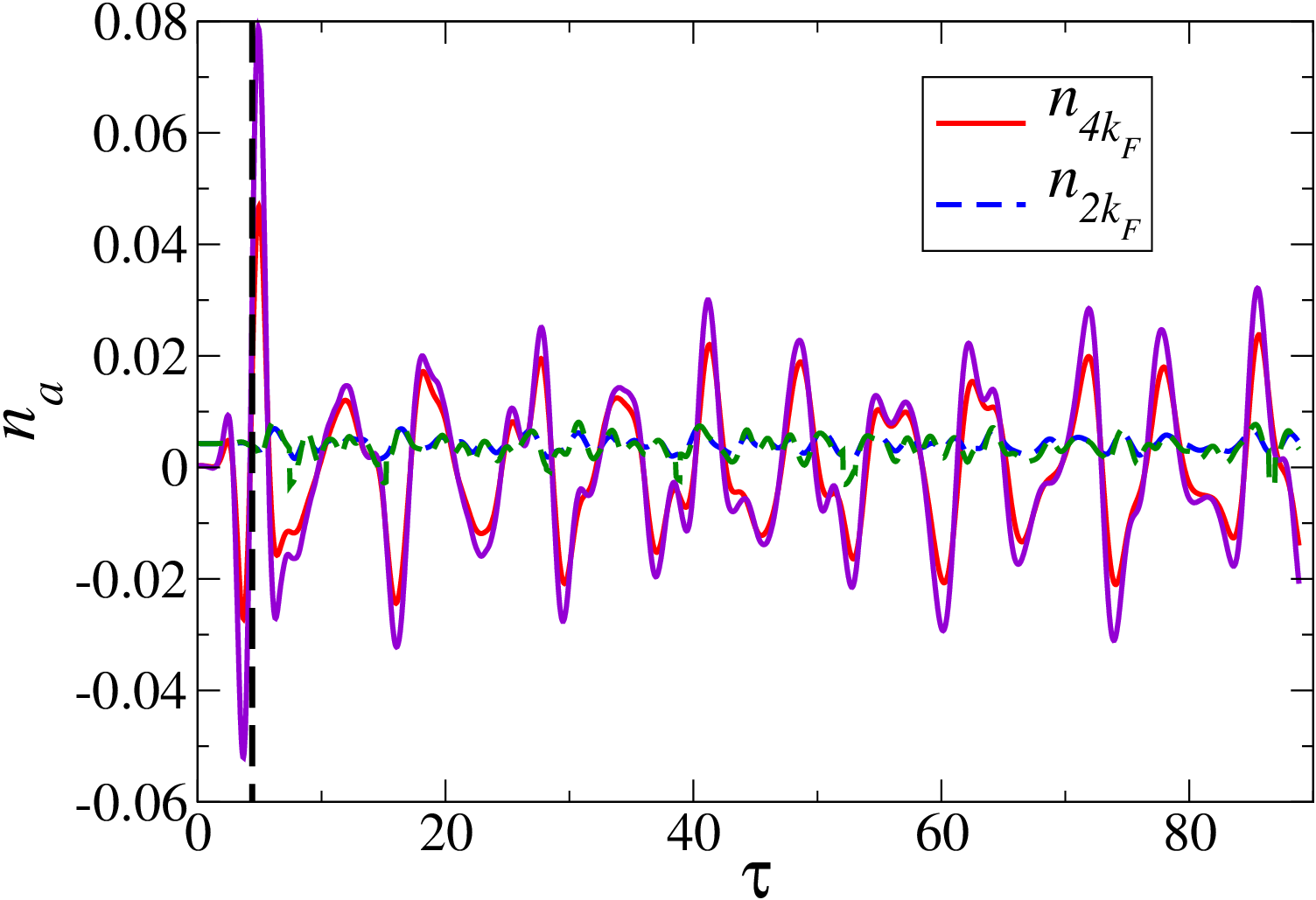}
\includegraphics*[width=.8\columnwidth]{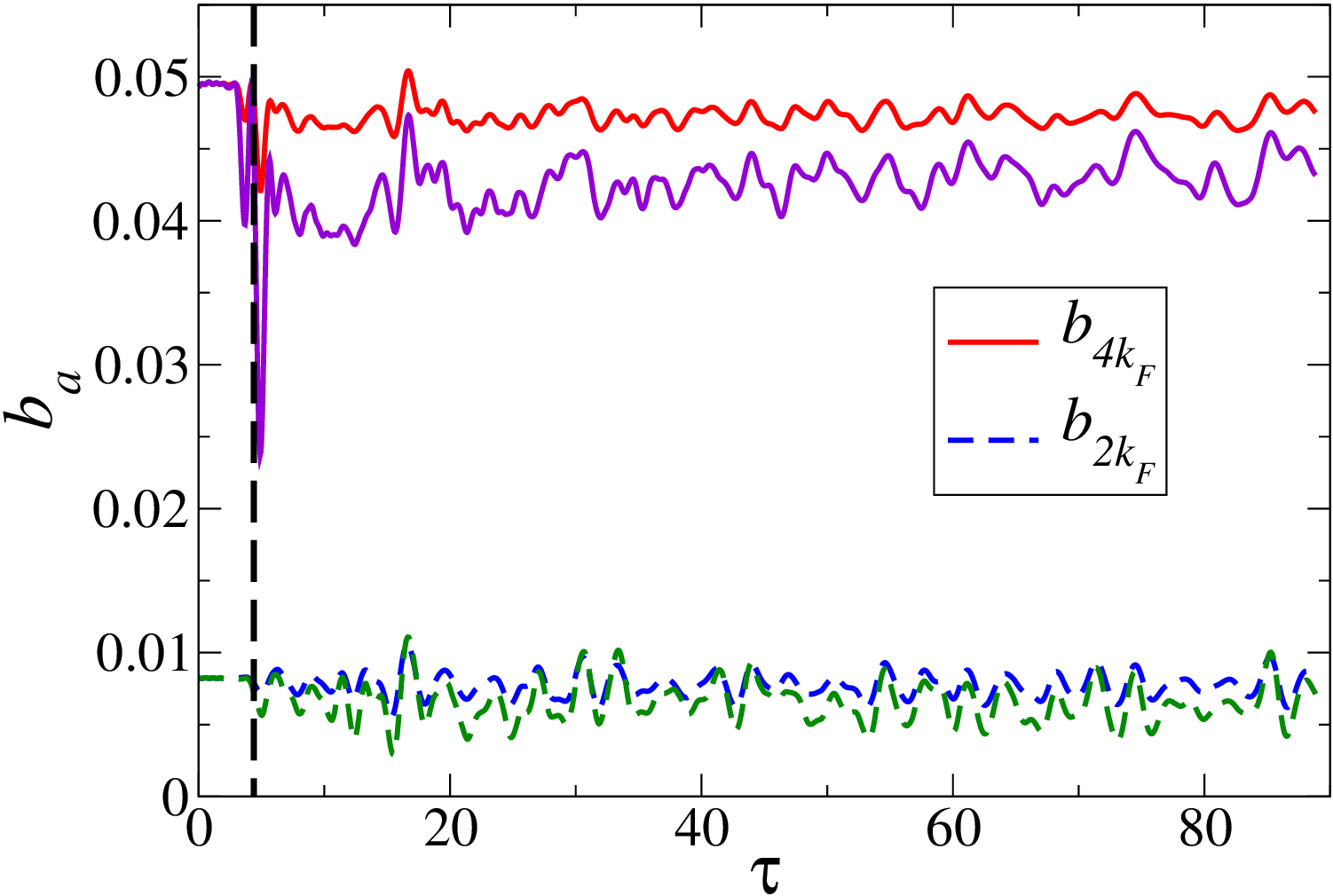}
\caption{Photoexcited time-evolution of the charge (top) and bond (bottom) instabilities, $a=2k_F$ and $4k_F$. The vertical dashed line corresponds to the maximum amplitude of the nonresonant light at $\tau_p$ and parameters: $\delta=0.35$, $U=8.24t$, $V=1.64t$, $A_0=5.5$ (red and blue) and 11 (violet and green), $\omega_{p}=2.5t$, $\sigma=1.1/t$.}
\label{fig:5}
\end{figure}

\begin{figure}
\includegraphics*[width=.8\columnwidth]{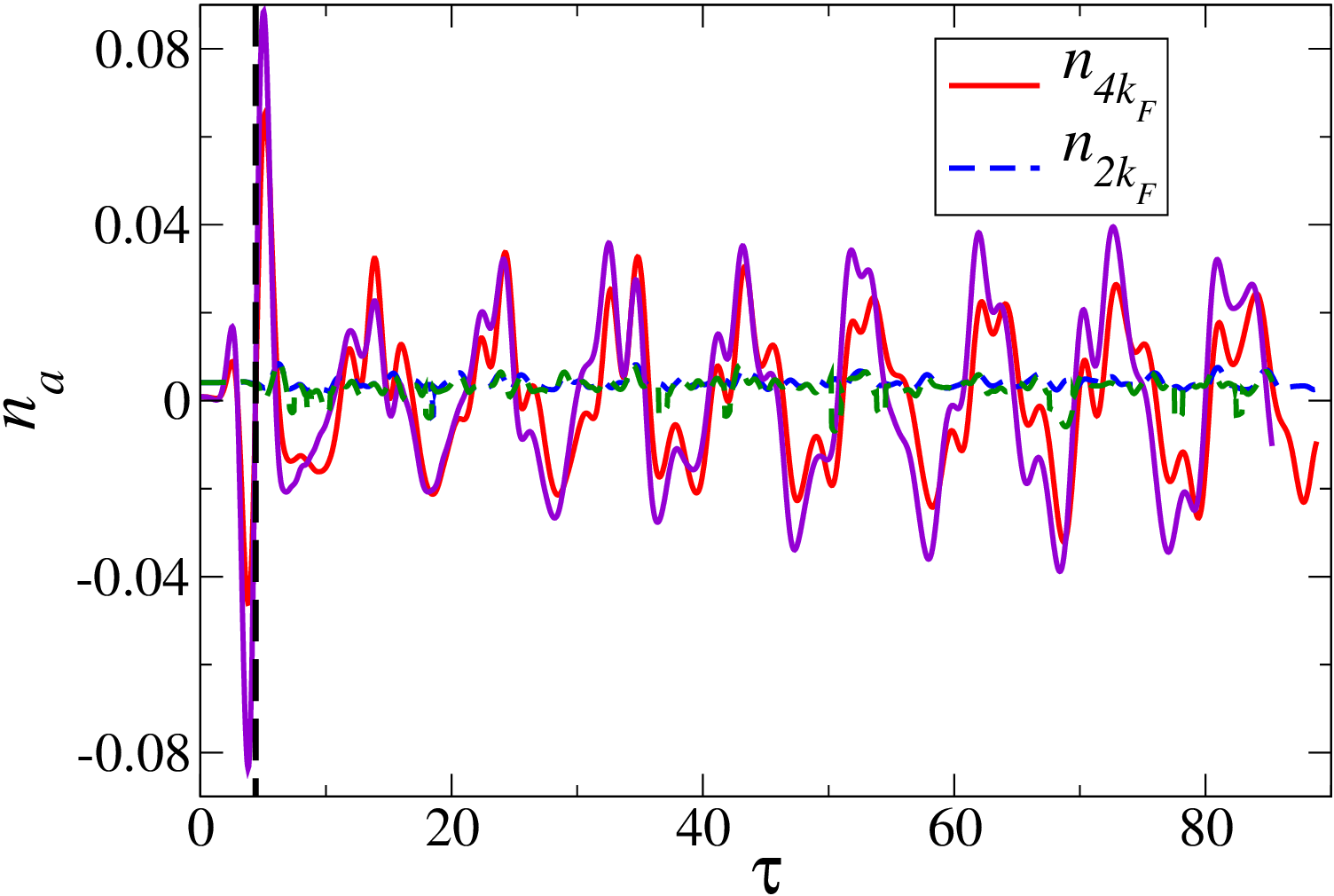}
\includegraphics*[width=.8\columnwidth]{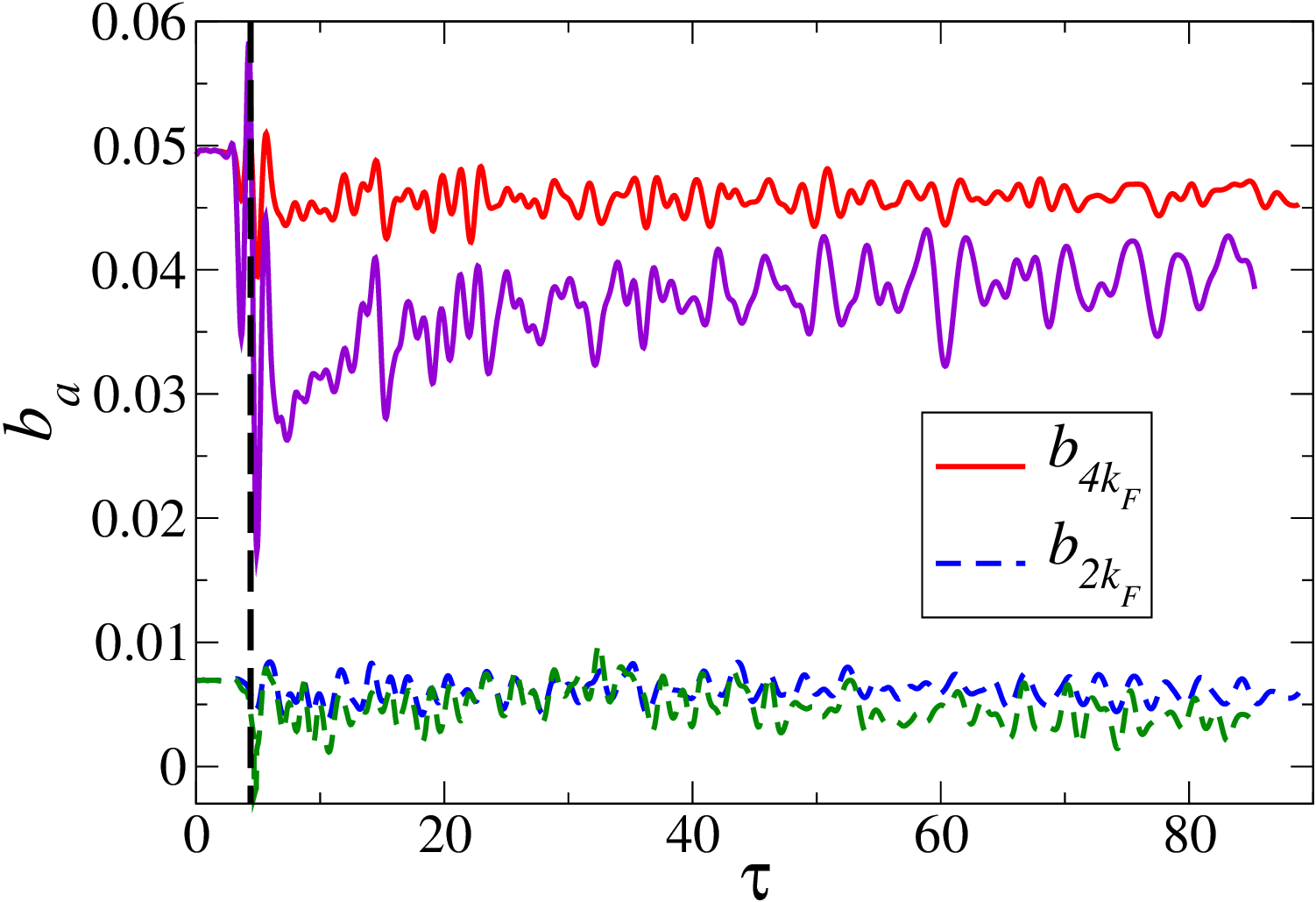}
\caption{Photoexcited time-evolution of the charge (top) and bond (bottom) instabilities, $a=2k_F$ and $4k_F$. The vertical dashed line corresponds to the maximum amplitude of the nonresonant light at $\tau_p$ and parameters: $\delta=0.35$, $U=8.24t$, $V=3.29t$, $A_0=7.5$ (red and blue) and 15 (violet and green), $\omega_{p}=2.5t$, $\sigma=1.1/t$.}
\label{fig:6}
\end{figure}

The photoinduced dynamics of the charge is shown in the top panel of Fig.~\ref{fig:4} for two different values of the amplitude of the incident radiation ($A_0=3.5$ and 7) and parameters $\delta=1.64$, $U=3.64t$, $V=0$. After the system is pumped, $n_{4k_F}$ oscillates with the same frequency as in the resonant case (see Fig.~\ref{fig:1}); i.e., intra- and interdimer excitations of energy $\omega=2t(1+\delta/2)$ dictate its time-dependent dynamics. The oscillations are accompanied by a decay in the amplitude as time evolves. The $n_{2k_F}$ instability, on the other hand, remains largely unaffected by the pulse in the time window explored. At later times there is a decay in the amplitude of the oscillations of the charge.

The nonresonant response of the bond instabilities, shown in Fig.~\ref{fig:4} (bottom panel), are also quite similar to the resonant case. Not only intra- and interdimer excitations, but also the low-energy absorption band of the spectrum ($\omega<2\Delta$) contribute to the photodynamics. Notice that in contrast with the charge case, the amplitude of the oscillations of $b_{4k_F}$ and $b_{2k_F}$ are differently affected by different amplitudes of the electric field, leading to different decaying rates for different values of $A_0$. Nevertheless, the frequency of the excitations are unaffected by the amplitude of the external radiation; this is a feature of nonresonant photodynamics.


Figure~\ref{fig:5} displays the dynamics of the electronic instabilities 
for the case $\delta=0.35$, $U=8.24t$, and $V=1.64t$, 
with amplitudes of the electric field $A_0=5.5$ and 11. In the case of 
charge instabilities, regardless of the intensity of the laser light, $n_{2k_F}$ is 
barely affected, contrary to $n_{4k_F}$ (Fig.~\ref{fig:5}, top). At time $\tau=\tau_{p}$, 
the $4k_F$ order is largely enhanced with a posterior reduction in the oscillation amplitude. 
A Fourier analysis (not shown here) revealed that the band of frequencies of the oscillations 
has a wider spectral range and smaller weights than in the resonant case (see Fig.~\ref{fig:2}). 
The characteristic energies associated with these frequencies are related with those of holon-antiholon pairs.

The time evolution of the bond order for the same parameters is shown in the bottom panel of Fig.~\ref{fig:5}. At time $\tau=\tau_{p}$, the $4k_F$ order is greatly reduced and then quickly restored (although not completely) followed by weak oscillations. On the other hand, $b_{2k_F}$ is only moderately affected by the pulse intensity around $\tau=\tau_{p}$; nevertheless, the field does create oscillations at later times similar to those of $b_{4k_F}$. The physical character of the oscillations is the same as for the charge instabilities.

The presence of excitons as the low-energy optical excitation is explored for the parameters $\delta=0.35$, $U=8.24t$, $V=3.29t$, with electric field amplitudes $A_0=7.5$ and 15. The behavior of the electronic instabilities does not change the system response to an intense light pulse around $\tau=\tau_{p}$ compared 
to the case shown in Fig.~\ref{fig:5}; however, the origin of the oscillations is quite different, 
as explained below. The time evolution of the charge and bond instabilities, shown in the top 
and bottom panels of Fig.~\ref{fig:6}, respectively, display similar behaviors to those in the case 
shown in Fig.~\ref{fig:3}. $n_{4k_F}$ increases as the pulse reaches its maximal value at 
$\tau=\tau_{p}$, and once the pulse is applied, $n_{4k_F}$ decreases and starts oscillating 
with a smaller amplitude. $n_{2k_F}$ is slightly affected in the time window considered. 
The nature of the oscillations can be attributed to the photogeneration of excitonic energy 
and the band of unbound pairs, just as in the case shown in Fig.~\ref{fig:3}.

The effect of the electric field on $b_{4k_F}$ and $b_{2k_F}$ is shown in the bottom 
panel of Fig.~\ref{fig:6}. At early times these instabilities are significantly reduced; 
this effect is stronger for the $4k_F$ instability. For times $\tau > t_{\rm pump}$, $b_{2k_F}$ 
quickly relaxes back to its original value at $\tau = 0$; this quick recovery is followed by 
clear oscillations. On the other hand, $b_{4k_F}$ displays a slower relaxation rate than 
$b_{2k_F}$ without reaching its original value. As in the case of the charge instability, 
the oscillations are associated with the energies of bound (excitons) and unbound 
(holon-antiholon pairs) optical excitations; the long-time physics of the photodynamics 
is again closely related to that of the resonant case.

\subsection{Discussion}
Assuming that the static phase diagram has the same states, although redistributed, 
as the nonstatic phase diagram, then interesting features appear. For instance, 
notice that when $n_{4k_F}=0$ for finite $\tau$, the system closely resembles the ground state ($\tau=0$); however, the values of $b_{4k_F}$ and $b_{2k_F}$ do not correspond to those of the ground state. Since we are studying a closed system, the electric field will pump energy into it, and it is expected that the long time behavior will be described by a thermal distribution.
However, 
it is possible to assign effective interaction parameters to these time-dependent 
states that are closely related to the original parameters in the phase diagram; 
in other words, one can envision each finite-$\tau$ state as a ground state 
of a Hamiltonian with parameters which depend on time. On the other hand, if we 
focus on the extreme values of $n_{4k_F}$, the effective parameters that we can 
assign to the time-dependent state will be farther away, in the phase diagram, 
from the original ones. This situation implies a vast sampling of the phase space through 
photoirradiation; whether this sampling is bounded or not remains as an open question. 
The above argument holds in general and does not depend on the parameters 
or the nature of the radiation.

The main difference between resonant and nonresonant pumping lies in how 
the system behaves around $\tau=\tau_{p}$ and the amplitude of the resulting 
oscillations. The general trend is that the charge instabilities, in the 
nonresonant case, will oscillate with a wider range of frequencies and 
lower amplitudes than in the resonant scenario, at least for $\tau >t_{\rm pump}$. 
For the bond instabilities, the system always shows a degree of relaxation after 
the photoexcitation and a considerable reduction of its value around $\tau=\tau_{p}$. 
In the resonant case, the photoinduced changes and oscillations of the instabilities 
are robust (the amplitude of the oscillations remains stable in the entire time window); 
whereas in the nonresonant case the electronic instabilities are greatly modified around 
$\tau=t_{\rm pump}$ followed by a still visible, albeit smaller, change in the amplitude 
of the oscillations.

This is in accord with the ponderomotive effect which comes into play when an off-resonant electric field is applied to the system~\cite{Wen08,Hirori11,Kampfrath13}. Indeed, there is a process which is not instantaneous and is inherent to the off-resonant response of the system to the 
accumulation of energy associated with the ponderomotive effect, 
that goes as $A_0^2$. After such energy has been absorbed, the system is capable 
of creating optical excitations displaying the consequent oscillatory behavior 
with the characteristic frequencies of those excitations. The observed delay, 
at times $\tau \gtrsim \tau_p$, in the creation of these excitations will depend 
on the energy associated with such excitations, $\omega$, and the amplitude of 
the electromagnetic pulse, $A_0$.

Notice also that the response of the charge order to different electric field 
amplitudes is different from that of the bond order. As expected from the 
ponderomotive effect, at earlier times $\tau\sim\tau_p$ both charge and bond instabilities 
have dynamics with similar frequencies but different amplitudes; however, for longer times, 
$\tau>t_{\rm pump}$, the charge excitations have similar amplitudes regardless 
of the value of $A_0$, whereas the bond instability clearly shows different 
amplitudes in its photodynamics for different values of $A_0$. The different 
dependence of the photodynamics on $A_0$, for charge and bond instabilities, 
remains in the way the energy accumulated is distributed along charge 
and bond excitations.


A general result from this study is that the 1D EPH model for charge-transfer systems, 
with no phonon dynamics, will tend to largely enhance the $4k_F$ charge instability and 
that this electronic instability fluctuates between (0101) and (1010) states. The 
time-dependent dynamics of the electronic instabilities shows the photoinduced melting 
of the $2k_F$ charge state and the concomitant replacement by the $4k_F$ ordered state. 
This ensuing dynamics is mediated by the photogeneration of optical excitations. On the 
other hand, the $2k_F$ bond and charge instabilities are not greatly affected by the 
presence of the electric field. We believe that the $4k_F$ instability is considerably 
more sensitive than the $2k_F$ instability to the excitation with light, probably due 
to the fact that the current operator, or equivalently the kinetic energy operator, 
only strongly couples $4k_F$ states (i.e., neighboring sites). We speculate that if 
one considers longer-range hopping, the $2k_F$ states will be largely affected by 
the time-dependent electric field. 

Regardless of the nature of the light interacting 
with the system, this scenario is observed in the photogenerated dynamics for all of 
the parameters studied in this work. Changing the parameters will only affect the 
characteristic frequencies, and at a minor level, the amplitude of the oscillations 
of the instabilities; in other words, the main role that the optical excitations (holon-antiholon pairs, excitons, etc.) play in the photodynamics is setting the frequency of the charge and bond oscillations.

The results presented in this work could have potential applications in the optoelectronics 
of charge-transfer systems. Taking advantage of the enormous positive/negative oscillating 
response of the $4k_F$ instability, organic charge-transfer systems can be used 
as switching devices with femtosecond response (provided that in such systems 
the vibrational degrees of freedom are slow compared to any other electronic 
time scale)~\cite{Chollet05,Onda08}.

\section{Conclusion\label{sec:con}}
In this publication, we have investigated the time-dependent behavior of the photoexcited 
electronic instabilities of a quarter-filled one-dimensional extended Peierls-Hubbard model, 
using static and time-dependent DMRG methods. Both charge and bond instabilities were studied 
as a function of time when the system is pumped with a resonant or nonresonant few-cycle 
electric field. Their resulting behavior can be explained in the light of the system optical 
excitations. Our main results show that the overall dynamics of the $4k_F$ bond and charge 
instabilities corresponds to a gigantic fluctuating behavior as a function of time. 
By contrast, the time-dependent response of the $2k_F$ instability (both for bond and charge) 
to the incident radiation displays a fairly smooth trend. These results remain valid whether 
the applied light pulse is in or off resonance with the optical excitations of the system, 
and regardless of the nature of such excitations. We argue that our calculations indicate 
that charge-transfer organic systems with slow phonon dynamics will display robust 
switching properties that can be potentially used in optoelectronic devices.

\acknowledgments
J.R.~acknowledges G.~B.~Martins for insightful conversations. Support by the Early Career Research Program, Scientific User Facilities Division, Basic Energy Sciences, US Department of Energy, under contract with UT-Battelle (J.R., K.A.) is acknowledged. A.E.F.~was supported by the National Science Foundation through grant DMR-1339564 and E.D.~through grant DMR-1404375.


\begin{thebibliography}{10}

\bibitem{Corkum07} P. B. Corkum and F. Krausz, \textit{Nat. Phys.} \textbf{3}, 381 (2007).

\bibitem{Goulielmakis07} E. Goulielmakis, V. S. Yakovlev, A. L. Cavalieri, M. Uiberacker, V. Pervak, A. Apolonski, R. Kienberger, U. Kleineberg, F. Krausz, \textit{Science} \textit{317}, 769 (2007).

\bibitem{Orenstein12} J. Orenstein, \textit{Physics Today} \textbf{65}, 44 (2012).

\bibitem{Kampfrath13} T. Kampfrath, K. Tanaka, and K. A. Nelson, \textit{Nature Photon.} \textbf{7}, 680 (2013).

\bibitem{Chollet05} M. Chollet, L. Guerin, Na. Uchida, S. Fukaya, H. Shimoda, T. Ishikawa, K. Matsuda, T. Hasegawa, A. Ota, H. Yamochi, G. Saito, R. Tazaki, S.-I. Adachi, S.-Y. Koshihara, \textit{Science} \textbf{307}, 86 (2005).

\bibitem{Onda08} K. Onda, S. Ogihara, K. Yonemitsu, N. Maeshima, T. Ishikawa, Y. Okimoto, X. Shao, Y. Nakano, H. Yamochi, G. Saito, and S.-Y. Koshihara, \textit{Phys. Rev. Lett.} \textbf{101}, 067403 (2008).

\bibitem{Benthien05} H. Benthien and E. Jeckelmann, \textit{Eur. Phys. J. B} \textbf{44}, 287 (2005).

\bibitem{Okamoto07} H. Okamoto, H. Matsuzaki, T. Wakabayashi, Y. Takahashi, and T. Hasegawa \textit{Phys. Rev. Lett.} \textbf{98}, 037401 (2007).

\bibitem{Maeshima05} N. Maeshima and K. Yonemitsu, \textit{J. Phys. Soc. Jpn.} \textbf{74}, 2671 (2005).

\bibitem{Maeshima06} N. Maeshima and K. Yonemitsu, \textit{Phys. Rev. B} \textbf{74}, 155105 (2006).

\bibitem{Maeshima07} N. Maeshima and K. Yonemitsu, \textit{J. Phys. Soc. Jpn.} \textbf{76}, 074713 (2007).

\bibitem{Yonemitsu07} K. Yonemitsu and N. Maeshima, \textit{Phys. Rev. B} \textbf{76}, 075105 (2007).

\bibitem{Yonemitsu09} K. Yonemitsu and N. Maeshima, Y. Tanaka, and S. Miyashita, \textit{J. Phys.: Conf. Ser.} \textbf{148}, 012054 (2009).

\bibitem{Lee09} J. D. Lee, \textit{Phys. Rev. B} \textbf{80}, 165101 (2009).

\bibitem{Moriya12} K. Moriya, N. Maeshima and K. I. Hino, \textit{Eur. Phys. J. B} \textbf{85}, 350 (2012).

\bibitem{Uemura12} H. Uemura, N. Maeshima, K. Yonemitsu, H. Okamoto, \textit{Phys. Rev. B} \textbf{85}, 125112 (2012).

\bibitem{Lu12} H. Lu, S. Sota, H. Matsueda, J. Bon\v ca, and T. Tohyama, \textit{Phys. Rev. Lett.} {\bf 109}, 197401 (2012),

\bibitem{Matsueda12} H. Matsueda, S. Sota, T. Tohyama, and S. Maekawa, \textit{J. Phys. Soc. Jpn.} {\bf 81}, 013701 (2012).

\bibitem{dmrg1} S. R. White, \textit{Phys. Rev. Lett.} {\bf 69}, 2863 (1992); \emph{ibid.}, \textit{Phys. Rev. B} {\bf 48}, 10345 (1993).

\bibitem{dmrg2} U. Schollw\"ock, \textit{Rev. Mod. Phys.} {\bf 77}, 259 (2005). 

\bibitem{dmrg3} K. Hallberg, \textit{Adv. Phys.} {\bf 55}, 477 (2006).

\bibitem{Ung94} K. C. Ung, S. Mazumdar, and D. Toussaint, \textit{Phys. Rev. Lett.} \textbf{73}, 2603 (1994).

\bibitem{Riera00} J. Riera and D. Poilblanc, \textit{Phys. Rev. B} \textbf{65}, 16243(R) (2000).

\bibitem{Shibata01} Y. Shibata, S. Nishimoto, and Y. Ohta, \textit{Phys. Rev. B} \textbf{64}, 235107 (2001).

\bibitem{Tsuchiizu01} M. Tsuchiizu, H. Yoshioka, and Y. Suzumura, \textit{J. Phys. Soc. Jpn.} \textbf{70}, 1460 (2001).

\bibitem{Kuwabara03} M. Kuwabara, H. Seo, and M. Ogata, \textit{J. Phys. Soc. Jpn.} \textbf{72}, 225 (2003).

\bibitem{Madsen02} L. B. Madsen, \textit{Phys. Rev. A} \textbf{65}, 053417 (2002).

\bibitem{Jeckelmann04} E. Jeckelmann, \textit{Phys. Rev. B} \textbf{67}, 075106 (2003).

\bibitem{White04} S. R. White and A. E. Feiguin, \textit{Phys. Rev. Lett.} \textbf{93}, 076401 (2004).

\bibitem{Daley04} A. J. Daley, C. Kollath, U. Schollw\"ock and G. Vidal, J. Stat. Mech.: Theor. Exp. \textbf{P04005} (2004).

\bibitem{Note1} The fittings have been performed in an $L/2$-site segment which is centered in the middle of the full $L$-site system to try to avoid Friedel oscillations introduced by the open boundaries. Some peaked behavior of the instabilities is observed due to convergence issues of the highly nonlinear fittings.

\bibitem{Note2} The energies obtained by a Fourier decomposition of the $\tau$-dependent instabilities were associated to the optical excitacions.

\bibitem{Wen08} H. Wen, M. Wiczer, and A. M. Lindenberg, \textit{Phys. Rev. B} \textbf{78}, 125203 (2008).

\bibitem{Hirori11} H. Hirori, K. Shinokita, M. Shirai, S. Tani, Y. Kadoya, and K. Tanaka, \textit{Nat. Comm.} \textbf{2}, 594 (2011).



\end{thebibliography}
\end{document}